\documentclass[12pt]{article}

\usepackage{amsmath, amssymb, amsfonts}
\usepackage{graphicx}
\usepackage{hyperref}
\usepackage{authblk}
\hypersetup{
    colorlinks=true,
    linkcolor=black,
    citecolor=black,
    filecolor=black,
    urlcolor=black,
    pdfborder={0 0 0}
}
\usepackage{geometry}
\usepackage[numbers,sort&compress]{natbib}
\usepackage{setspace}
\usepackage{titlesec}
\usepackage{caption}
\usepackage{float}
\usepackage{booktabs}
\usepackage{adjustbox}
\usepackage{array}
\usepackage{threeparttable}
\usepackage{arydshln}
\usepackage{multirow}
\usepackage{times}
\usepackage{dsfont}   

\titleformat{\section}
  {\normalfont\Large\bfseries}{\thesection}{1em}{}
\titleformat{\subsection}
  {\normalfont\large\bfseries}{\thesubsection}{1em}{}

\title{Bayesian Semiparametric Joint Dynamic Model for Multitype Recurrent Events and a Terminal Event}

\author{
\begin{flushleft}
\textbf{Mithun Kumar Acharjee}\textsuperscript{1,*} \quad \textbar \quad
\textbf{AKM Fazlur Rahman}\textsuperscript{1}

\vspace{1.2em}

\textsuperscript{1}Department of Biostatistics, School of Public Health, University of Alabama at Birmingham, Birmimngham, AL 35294, USA.

\vspace{1.2em}

*Corresponding author:

Mithun Kumar Acharjee, Email: \texttt{acharjee@uab.edu}
\end{flushleft}
}

\date{}

\begin{document}

\maketitle


\begin{abstract}
In many biomedical research, recurrent events such as myocardial infraction, stroke, and heart failure often result in  a terminal outcome such as death. Understanding the relationship among the multi-type recurrent events and terminal event is essential for developing interventions  to prolong the  terminal event such as death. This study introduces a Bayesian semiparametric joint dynamic model for type-specific hazards that quantifies how the type-specific event history dynamically changes the intensities of each recurrent event type and the terminal event over calendar time. The framework jointly captures unmeasured heterogeneity through a shared frailty term, cumulative effects of past recurrent events on themselves and terminal events, and the effects of covariates. Gamma process priors (GPP) are used as a nonparametric prior for the baseline cumulative hazard function (CHF) and parametric priors for covariates and frailty. For a more accurate risk assessment, this model provides an analytical closed-form estimator of cumulative hazard functions (CHF) and frailties. The Breslow-Aalen-type estimators of CHFs are special cases of our estimators when the precision parameters are set to zero. We evaluate the performance of the model through extensive simulations and apply the method to the Antihypertensive and Lipid-Lowering Treatment to Prevent Heart Attack Trial (ALLHAT). The analysis offers a practical past event effect based risk assessment for acute and chronic cardiovascular recurrent events with a terminal end point death and provides new information to support the prevention and treatment of cardiovascular disease to clinicians.
\end{abstract}

\section{Introduction}

Cardiovascular disease (CVD) is the leading cause of death worldwide \cite{WHO2024CVD}. CVD patients often experience more than one event. Over the years, many have had multiple types of events (recurrent or non-recurrent), such as myocardial infarction, stroke, and congestive heart failure, before experiencing a terminal event (e.g., death). Classic methods for recurrent events on the calendar-time scale, including the Andersen–Gill formulation and the Prentice–Williams–Peterson family, provide flexible regressions with clear clinical contrasts. However, they primarily target recurrences of a single event type, rather than multiple types \cite{AndersenGill1982, PWP1981}. For joint analyses with death, shared-frailty models connect the recurrent and terminal processes through a random effect, capturing unmeasured heterogeneity while keeping baseline hazards flexible. These tools are well documented and supported by software such as \texttt{frailtypack} \cite{Liu2004, Rondeau2007, RondeauJSS2012, CookLawless2002, AmorimCai2015}. In practice, clinical data often include several types of recurrent events plus a terminal event. However, the effect of past events on future risk is usually modeled only indirectly (e.g., through a frailty term) or in separate components, rather than with a clear, time-updated history term on the calendar-time scale.

Building on single-type foundations, several lines of work now incorporate dynamic modeling of past-event impact to address multitype recurrent events with death, which we acknowledge and distinguish from our contribution. One line of research uses multistate models in which biomarker processes carry “past-event feedback,” creating complex links between latent biology and event risks \cite{MaDaiPan2022, Rizopoulos2011}. In these models, past events influence risk primarily through the biomarkers, rather than through a direct, time-updated history vector that updates all recurrent intensities and the terminal hazard over calendar time \cite{MaDaiPan2022, Rizopoulos2011}. A second line develops dynamic risk models for multitype recurrent and terminal events. Liu and Pe{\~n}a \cite{LiuPena2015} proposed a semiparametric frequentist dynamic model for recurrent competing risks and a terminal event, fit by the EM algorithm on the calendar-time scale. In simulations, the frailty estimate showed sizable bias, consistent with the well-known sensitivity of EM to starting values \cite{McLachlanKrishnan2008, IbrahimChenSinha2001, Acharjee2025}. A third line—joint models with internal covariates or semiparametric transformation models—does allow the number of prior events to enter, but typically not as a time-updated past event history vector on the calendar-time scale that simultaneously drives all recurrent hazards and the terminal hazard within a unified Bayesian semiparametric joint framework \cite{ZengLin2009, HuangWang2004, Rondeau2007}.

We address this gap with a Bayesian semiparametric joint dynamic model for multitype recurrent events with a terminal event on the calendar-time scale. We represent a patient's event history as a time-updated summary that directly modifies risks as events accumulate. As different event types occur over time, these evolving histories enter explicitly into the intensities for each recurrent event type and for death, while a shared frailty accounts for unmeasured heterogeneity.  Concretely, the model estimates dynamic past-event effects of three kinds: (i) same-type effects, which quantify how past events of a given type alter future risk of that same type; (ii) cross-type effects, which quantify how the history of one type shifts the risks for other recurrent types; and (iii) terminal effects, which quantify how the accumulated multitype history influences the hazard of death. These yield direct, interpretable past-event effects on both recurrent and terminal processes—not merely indirect dependence through a random effect \cite{Liu2004}. We also derive a closed-form estimator for the frailty term, providing stable and efficient updates.

To retain flexibility without heavy tuning, we assign gamma-process priors on the cumulative baseline hazards. This approach in Bayesian survival analysis links naturally to cumulative-hazard estimators used with Cox-type models and yields simple conditional updates in MCMC \cite{Kalbfleisch1978, DykstraLaud1981, Hjort1990, Lin2007, IbrahimChenSinha2001}. In practice, it provides regularization and interpretability while avoiding knot selection and large matrix operations. This yields a simple, robust MCMC algorithm with low computation time compared with high-state multistate models or spline-based baselines \cite{IbrahimChenSinha2001}.

\begin{figure}[!ht]
\centering
\includegraphics[width=\textwidth]{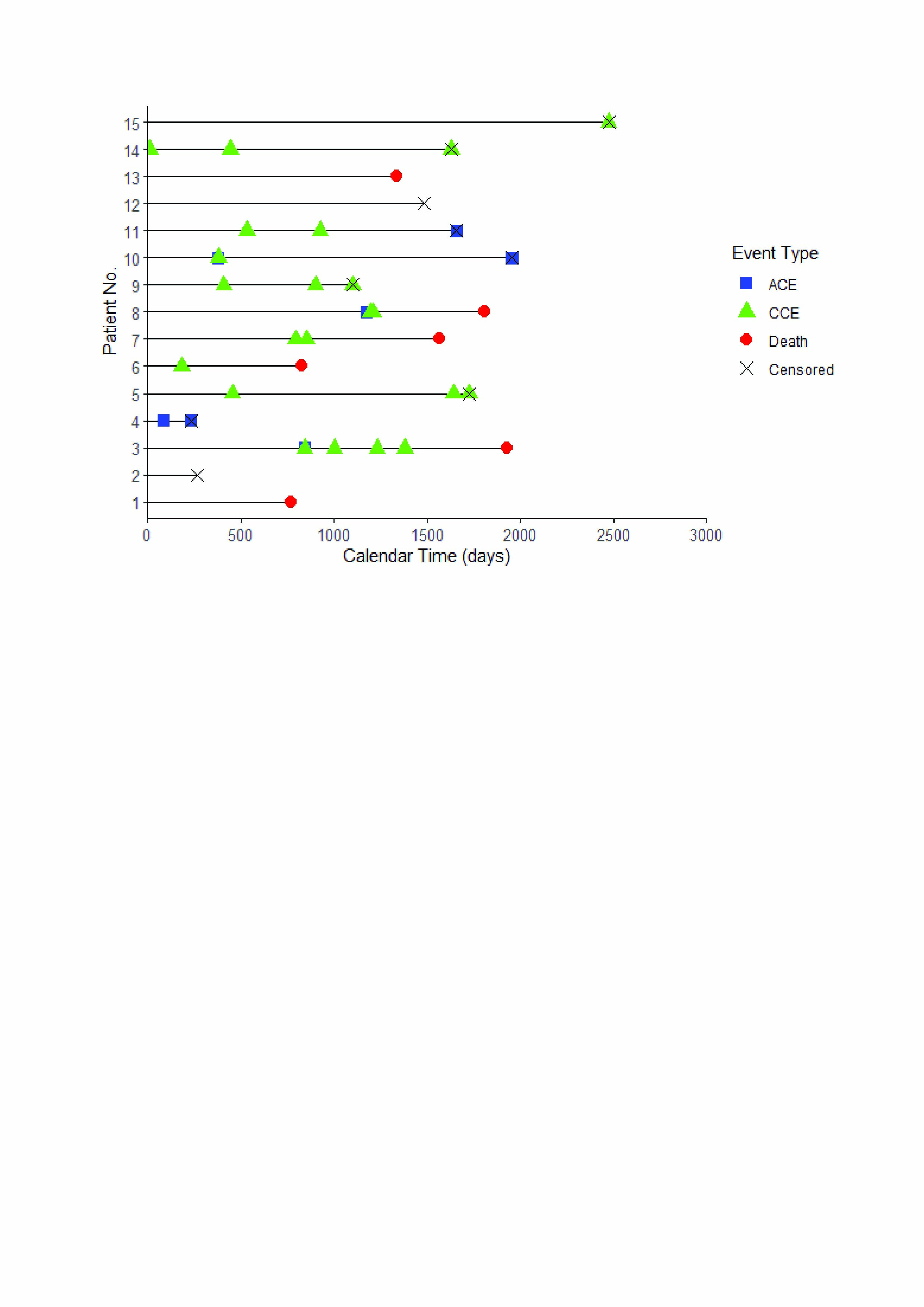}
\vspace{-13.5cm}
\caption{Multitype recurrent events and a terminal event for 15 selected participants from the ALLHAT trial.}
\label{fig:Figure_1.jpg}
\end{figure}

A simple case illustrates the clinical importance. In the ALLHAT hypertension trial (Figure \ref{fig:Figure_1.jpg}), participants accrued acute events (myocardial infarction, stroke) and chronic complications (congestive heart failure, peripheral arterial disease) over years and some died. Clinicians want to know, given the history up to today, how prior events—by type—change the risk of the next type of event and the risk of death in each randomized treatment. Our model provides those estimates directly on the calendar-time scale while adjusting for unmeasured heterogeneity through frailty \cite{ALLHAT2002}.

Finally, the paper is organized as follows. Section~\ref{sec:method} presents the proposed Bayesian semiparametric joint dynamic model, including formulation on the calendar-time scale, the construction of the rho function (impact of a function expressing the past event impact to the future risk), the role of shared frailty, and the use of priors of the gamma-process for cumulative baseline hazards. It also derives the likelihood and conditional posterior distributions and describes the MCMC algorithm. Section~\ref{sec:sim} reports results from simulation studies that evaluate the accuracy and robustness of the model estimation under different data-generating scenarios. Section~\ref{sec:app} applies the proposed model to the ALLHAT hypertension trial, demonstrating how the estimated effects of the multitype history clarify downstream risks and improve clinical interpretation. Section~\ref{sec:conc} concludes with a discussion of the main findings, the methodological implications, and the potential directions for future research.

\section{Methodology}\label{sec:method}

\subsection{Joint Dynamic Model}

Consider $n$ independent subjects indexed by $i=1,2,\ldots,n$, each associated with a random  component also know as frailty $W_i$ that could  effect recurrent event and terminal event occurrences  for a give subject. We assume $W_i \stackrel{\text{i.i.d.}}{\sim} \mathrm{Gamma}(\nu,\nu)$ with $\mathbb{E}(W_i)=1$ and $\mathrm{Var}(W_i)=1/\nu$. Let $\mathbf{X}_i=(X_{i1},X_{i2},\ldots,X_{ip})^\top$ denote a $p$-dimensional covariate vector observed for subject $i$. Each subject is followed over the interval $[0,\tau_i]$, where $\tau_i$ is the administrative censoring time. For $Q$ types of recurrent events, let $N_{qi}^\dagger(t)$ denote the counting process for the $q$th recurrent event type of subject $i$, defined as $N_{qi}^\dagger(t)=\sum_{j=1}^\infty \mathds{1}\{S_{qij}\le t,\, S_{qij}\le \min(\tau_i,T_{0i})\}$, where $S_{qij}$ represents the calendar time of the $j$th occurrence of type-$q$ event for subject $i$ and $T_{0i}$ is the terminal event time. The vector of past event occurrences up to time $t$ is denoted by $\boldsymbol{N}_i^{\dagger}(t^-)^\top=(N_{1i}^\dagger(t^-),N_{2i}^\dagger(t^-),\ldots,N_{Qi}^\dagger(t^-))$, which summarizes the complete multitype recurrent event history for subject $i$.

Let $\lambda_{0q}(t)$ and $\lambda_{00}(t)$ denote the baseline hazard functions for the type-q recurrent event and the terminal event, respectively. The regression coefficients $\beta_q$ and $\beta_0$ describe the effects of covariates on the hazards of the type-q recurrent event and terminal event processes. To account for the effect of prior event occurrences on the current event intensity, we define dynamic history functions $\rho_q[\boldsymbol{N}_i^{\dagger}(t^-);\boldsymbol{\alpha}_q]$ for the type-q recurrent event and $\rho_0[\boldsymbol{N}_i^{\dagger}(t^-);\boldsymbol{\gamma}]$ for the terminal event, where $\boldsymbol{\alpha}_q=(\alpha_{q1},\alpha_{q2},\ldots,\alpha_{qQ})^\top$ and $\boldsymbol{\gamma}=(\gamma_1,\gamma_2,\ldots,\gamma_Q)^\top$ are vectors of nonnegative parameters quantifying how past occurrences of each event type influence the risk of future events. One example of the rho function is 
\[
\rho_q[\boldsymbol{N}_i^{\dagger}(t^-);\boldsymbol{\alpha}_q]=1+\boldsymbol{N}_i^{\dagger}(t^-)^\top\boldsymbol{\alpha}_q, \qquad
\rho_0[\boldsymbol{N}_i^{\dagger}(t^-);\boldsymbol{\gamma}]=1+\boldsymbol{N}_i^{\dagger}(t^-)^\top\boldsymbol{\gamma},
\]
where the special case $\boldsymbol{\alpha}_q=\mathbf{0}$ or $\boldsymbol{\gamma}=\mathbf{0}$ corresponds to a model with no past event history. 

Conditional on $(W_i,\mathbf{X}_i)$ and the past event history $\mathcal{H}_i(t^-)=\{\boldsymbol{N}_i^{\dagger}(u):0<u<t\}$, the joint dynamic intensities of the type-$q$ recurrent events and the terminal event for subject $i$ are specified as
\begin{equation}
\label{eqn1}
\left\{
\begin{aligned}
\lambda_{qi}\!\left(t\,\middle|\,W_i,\mathbf{X}_i\right)
&= \lambda_{0q}(t)\;W_i\;\exp\!\big(\beta_q^\top\mathbf{X}_i\big)\;
\rho_q\!\big[\boldsymbol{N}_i^{\dagger}(t^-);\boldsymbol{\alpha}_q\big],\quad q=1,\ldots,Q,\\[2mm]
\lambda_{0i}\!\left(t\,\middle|\,W_i,\mathbf{X}_i\right)
&= \lambda_{00}(t)\;W_i\;\exp\!\big(\beta_0^\top\mathbf{X}_i\big)\;
\rho_0\!\big[\boldsymbol{N}_i^{\dagger}(t^-);\boldsymbol{\gamma}\big].
\end{aligned}
\right.
\end{equation}
where $\lambda_{qi}(t)$ and $\lambda_{0i}(t)$ denote the subject-specific intensities given the frailty, covariates, and event history. This formulation links all recurrent and terminal processes through the shared frailty $W_i$, the covariate effects $\beta_q$ and $\beta_0$, and the dynamic history functions $\rho_q(\cdot)$ and $\rho_0(\cdot)$, yielding a comprehensive semiparametric framework for modeling multitype recurrent and terminal events. For each subject $i$, the number of observed events for the type-q recurrent process and the terminal event are given by
\[
K_{qi}=\max\{k\ge0:S_{qik}\le\min(\tau_i,T_{0i})\}, \qquad
K_{0i}=\mathds{1}\{T_{0i}\le\tau_i\},
\]
and the observable random vector (data) is
\[
D_{i}=\big(\tau_i,\,T_{0i},\,K_{qi},\,K_{0i},\,T_{qi1},\,T_{qi2},\,\ldots,\,T_{qiK_{qi}},\,\min(\tau_i,T_{0i})-S_{qiK_{qi}}\big).
\]

\subsection{Likelihood}

We construct the likelihood using the counting-process formulation of Jacod \cite{jacod1975multivariate} and Andersen et al. \cite{andersen2012statistical}.  
For each subject $i$ and event type $q$, define the counting process
\[
N_{qi}^\dagger(t)=\sum_{j=1}^{\infty}\mathds{1}\{S_{qij}\le t,\, S_{qij}\le \min(\tau_i,T_{0i})\},
\]
where $S_{qij}$ is the calendar time of the $j$th type-$q$ recurrence, and define its increment over $(t,t+\Delta t]$ as $N_{qi}^\dagger(\Delta t)=N_{qi}^\dagger(t+\Delta t)-N_{qi}^\dagger(t)$.  
The corresponding at-risk indicator is
\[
Y_i^\dagger(t)=\mathds{1}\{\min(\tau_i,T_{0i})\ge t\},
\]
which equals 1 when subject $i$ is under observation and event-free at time $t$.

Given $(W_i,\mathbf{X}_i)$ and the event history $\mathcal{H}_i(t^-)$, the conditional intensity functions for $N_{qi}^\dagger(t)$ and $N_{0i}^\dagger(t)$ are specified in Eq.~(\ref{eqn1}).  
Then, by the general theory of multitype counting processes, the conditional likelihood for the observed data over $[0,s]$ is
\begin{align}
L(s\mid \boldsymbol{D},\mathbf{W})
&= \prod_{i=1}^{n}
\Bigg\{
\prod_{q=1}^{Q}
\Bigg[
\Bigg(\prod_{t\in[0,s]}
[\lambda_{qi}(t\mid W_i,\mathbf{X}_i)]^{N_{qi}^\dagger(\Delta t)}\Bigg)
\exp\!\Bigg[-\int_0^s Y_i^\dagger(u)\,\lambda_{qi}(u\mid W_i,\mathbf{X}_i)\,du\Bigg]
\Bigg]
\nonumber\\[-2pt]
&\qquad\times
\Bigg[
\Bigg(\prod_{t\in[0,s]}
[\lambda_{0i}(t\mid W_i,\mathbf{X}_i)]^{N_{0i}^\dagger(\Delta t)}\Bigg)
\exp\!\Bigg[-\int_0^s Y_i^\dagger(u)\,\lambda_{0i}(u\mid W_i,\mathbf{X}_i)\,du\Bigg]
\Bigg]
\Bigg\},
\label{eq2}
\end{align}
where $\mathbf{W}=(W_1,\ldots,W_n)$ and $\boldsymbol{D}=\{D_i,\mathbf{X}_i\}_{i=1}^n$ denote the observed data and covariates.

Substituting the dynamic intensities from Eq.~(\ref{eqn1}) into \eqref{eq2} yields
\begin{align}
L(s\mid \boldsymbol{D},\mathbf{W})
&=\prod_{i=1}^{n}
\Bigg\{
\prod_{q=1}^{Q}
\Bigg[
\Bigg(\prod_{t\in[0,s]}
[W_i\,\exp(\mathbf{X}_i^\top\beta_q)\,
Y_i^\dagger(t)\,
\rho_q[\boldsymbol{N}_i^\dagger(t^-);\boldsymbol{\alpha}_q]\,
\lambda_{0q}(t)]^{N_{qi}^\dagger(\Delta t)}\Bigg)
\nonumber\\[-2pt]
&\qquad\qquad\times
\exp\!\Bigg[-W_i\,\exp(\mathbf{X}_i^\top\beta_q)
\int_0^s Y_i^\dagger(u)\,
\rho_q[\boldsymbol{N}_i^\dagger(u^-);\boldsymbol{\alpha}_q]\,
\lambda_{0q}(u)\,du\Bigg]
\Bigg]
\nonumber\\
&\qquad\times
\Bigg[
\Bigg(\prod_{t\in[0,s]}
[W_i\,\exp(\mathbf{X}_i^\top\beta_0)\,
Y_i^\dagger(t)\,
\rho_0[\boldsymbol{N}_i^\dagger(t^-);\boldsymbol{\gamma}]\,
\lambda_{00}(t)]^{N_{0i}^\dagger(\Delta t)}\Bigg)
\nonumber\\[-2pt]
&\qquad\qquad\times
\exp\!\Bigg[-W_i\,\exp(\mathbf{X}_i^\top\beta_0)
\int_0^s Y_i^\dagger(u)\,
\rho_0[\boldsymbol{N}_i^\dagger(u^-);\boldsymbol{\gamma}]\,
\lambda_{00}(u)\,du\Bigg]
\Bigg]
\Bigg\}.
\nonumber
\end{align}

For computational implementation, let $t_{(1)}<t_{(2)}<\cdots<t_{(M)}$ denote the ordered distinct event times with $t_{(0)}=0$.  
Define the cumulative hazard increments $\Lambda_{0q}(\Delta t_{(j)})=\Lambda_{0q}(t_{(j)})-\Lambda_{0q}(t_{(j-1)})$ and $\Lambda_{00}(\Delta t_{(j)})=\Lambda_{00}(t_{(j)})-\Lambda_{00}(t_{(j-1)})$.  
The discrete representation of the likelihood is then
\begin{align}
&L(\Lambda_{0q}(\cdot),\Lambda_{00}(\cdot),\beta_q,\beta_0,\nu\mid \boldsymbol{D},\mathbf{W})
\nonumber\\
&\quad=
\prod_{i=1}^{n}
\Bigg\{
\prod_{q=1}^{Q}
\Bigg[
\Bigg(\prod_{j=1}^{M}
[W_i\,\exp(\mathbf{X}_i^\top\beta_q)\,
Y_i^\dagger(t_{(j)})\,
\rho_q[\boldsymbol{N}_i^\dagger(t_{(j)}^-);\boldsymbol{\alpha}_q]\,
\Lambda_{0q}(\Delta t_{(j)})]^{N_{qi}^\dagger(\Delta t_{(j)})}\Bigg)
\nonumber\\[-2pt]
&\qquad\qquad\times
\exp\!\left[-W_i\,\exp(\mathbf{X}_i^\top\beta_q)
\sum_{j=1}^{M}
Y_i^\dagger(t_{(j)})\,
\rho_q[\boldsymbol{N}_i^\dagger(t_{(j)}^-);\boldsymbol{\alpha}_q]\,
\Lambda_{0q}(\Delta t_{(j)})\right]
\Bigg]
\nonumber\\
&\qquad\times
\Bigg[
\Bigg(\prod_{j=1}^{M}
[W_i\,\exp(\mathbf{X}_i^\top\beta_0)\,
Y_i^\dagger(t_{(j)})\,
\rho_0[\boldsymbol{N}_i^\dagger(t_{(j)}^-);\boldsymbol{\gamma}]\,
\Lambda_{00}(\Delta t_{(j)})]^{N_{0i}^\dagger(\Delta t_{(j)})}\Bigg)
\nonumber\\[-2pt]
&\qquad\qquad\times
\exp\!\left[-W_i\,\exp(\mathbf{X}_i^\top\beta_0)
\sum_{j=1}^{M}
Y_i^\dagger(t_{(j)})\,
\rho_0[\boldsymbol{N}_i^\dagger(t_{(j)}^-);\boldsymbol{\gamma}]\,
\Lambda_{00}(\Delta t_{(j)})\right]
\Bigg]
\Bigg\},
\label{eq3}
\end{align}
which serves as the basis for Bayesian estimation under gamma-process priors on $\Lambda_{0q}(\cdot)$ and $\Lambda_{00}(\cdot)$.

\subsection{Prior Specifications and Conditional Posteriors}
The unknown parameters of interest are \(\Lambda_{0q}(\cdot), \Lambda_{00}(\cdot), \nu, \beta_q, \beta_0, \boldsymbol{\alpha}_q, \boldsymbol{\gamma}\); we also derive the conditional distribution of the unobservable frailty vector \(W_1, W_2, \dots, W_n\), which will be involved in the Bayes estimators of \(\Lambda_{0q}(\cdot)\) and \(\Lambda_{00}(\cdot)\). Following Kalbfleisch \cite{Kalbfleisch1978}, we assume \(\Lambda_{0q}(\cdot)\) and \(\Lambda_{00}(\cdot)\) have nonparametric gamma-process priors \(\Lambda_{0q}(\cdot) \sim \mathcal{G}_{c_q,\Lambda_{0q}^*(\cdot)}\) and \(\Lambda_{00}(\cdot) \sim \mathcal{G}_{c_0,\Lambda_{00}^*(\cdot)}\), where \(\Lambda_{0q}^*(\cdot)\) and \(\Lambda_{00}^*(\cdot)\) are completely known mean intensity functions for type-\(q\) recurrent events and the terminal event, respectively. Also, \(c_q\) and \(c_0\) represent the precision of the prior measure for recurrent and terminal events, respectively. Then, for the ordered distinct event times \(t_{(1)}<\cdots<t_{(M)}\) with \(t_{(0)}=0\),
\[
\begin{aligned}
\Lambda_{0q}\!\big(\Delta t_{(j)}\big) 
&\sim \mathrm{Gamma}\!\left(c_q\,\Lambda_{0q}^*\!\big(\Delta t_{(j)}\big),\, c_q\right), 
&& j=1,\ldots,M, \\[6pt]
\Lambda_{00}\!\big(\Delta t_{(j)}\big) 
&\sim \mathrm{Gamma}\!\left(c_0\,\Lambda_{00}^*\!\big(\Delta t_{(j)}\big),\, c_0\right), 
&& j=1,\ldots,M.
\end{aligned}
\] where \(\Lambda_{0q}(\Delta t_{(j)})=\Lambda_{0q}(t_{(j)})-\Lambda_{0q}(t_{(j-1)})\) and \(\Lambda_{00}(\Delta t_{(j)})=\Lambda_{00}(t_{(j)})-\Lambda_{00}(t_{(j-1)})\).

\noindent We use the notation \(\pi (\Lambda_{0q}(\Delta t_{(j)}))\) to denote the prior distribution of \(\Lambda_{0q}(\Delta t_{(j)})\) and similarly for \(\Lambda_{00}(\Delta t_{(j)})\). That is,
\begin{equation*}
\begin{aligned}
    \pi(\Lambda_{0q}(\Delta t_{(j)})) & \sim  \mathrm{Gamma}\!\big(c_q\Lambda_{0q}^*(\Delta t_{(j)}),\,c_q\big), \quad j = 1,\dots,M, \\
    \pi(\Lambda_{00}(\Delta t_{(j)})) & \sim  \mathrm{Gamma}\!\big(c_0\Lambda_{00}^*(\Delta t_{(j)}),\,c_0\big), \quad j = 1,\dots,M.
\end{aligned}
\end{equation*}

\noindent We assume the prior of \(\nu\) is Gamma with known shape \(\zeta\) and rate \(\eta\). The priors for \(\beta_q\) and \(\beta_0\) are specified as \(p\)-dimensional multivariate normal distributions with known mean vectors \(\mu_{\beta_q}\) and \(\mu_{\beta_0}\), and corresponding variance–covariance matrices \(\Sigma_{\beta_q}\) and \(\Sigma_{\beta_0}\), respectively. For the dynamic history parameters, we place independent Gamma priors componentwise on \(\boldsymbol{\alpha}_q=(\alpha_{q1},\ldots,\alpha_{qQ})^\top\) and \(\boldsymbol{\gamma}=(\gamma_1,\ldots,\gamma_Q)^\top\):
\begin{equation*}
\begin{aligned}
    \pi(\nu) & \sim \mathrm{Gamma}(\zeta, \eta),\\
    \pi(\beta_q) & \sim N_p(\mu_{\beta_q}, \Sigma_{\beta_q}),\\
    \pi(\beta_0) & \sim N_p(\mu_{\beta_0}, \Sigma_{\beta_0}),\\
    \pi(\boldsymbol{\alpha}_q) &\;\propto\; \prod_{\ell=1}^Q \alpha_{q\ell}^{\,a_{q\ell}-1}\exp(-b_{q\ell}\alpha_{q\ell}),\\
    \pi(\boldsymbol{\gamma}) &\;\propto\; \prod_{\ell=1}^Q \gamma_{\ell}^{\,c_{\ell}-1}\exp(-d_{\ell}\gamma_{\ell}).
\end{aligned}
\end{equation*}
Using Eq.~(\ref{eq3}) and the above-specified priors, we define the joint posterior distribution of \\ \(\{\Lambda_{0q}(\cdot), \Lambda_{00}(\cdot), \nu, \beta_q, \beta_0, \boldsymbol{\alpha}_q, \boldsymbol{\gamma}\}\) via
\begin{align}
    p(\Lambda_{0q}(\cdot),\Lambda_{00}(\cdot), \mathbf{W}, \nu, \beta_q, \beta_0, \boldsymbol{\alpha}_q, \boldsymbol{\gamma} \mid \boldsymbol{D}) &\propto L(\Lambda_{0q}(\cdot), \Lambda_{00}(\cdot), \beta_q, \beta_0, \nu \mid \boldsymbol{D}, \mathbf{W}) \nonumber\\
    &\mathrel{\times} \pi(\Lambda_{0q}(\cdot))\,\pi(\Lambda_{00}(\cdot))\,\pi(\nu)\,\pi(\beta_q)\,\pi(\beta_0)\,\pi(\boldsymbol{\alpha}_q)\,\pi(\boldsymbol{\gamma}),
\label{eq4}
\end{align}
where \(\mathbf{W}=(W_1,\ldots,W_n)\) and \(\boldsymbol{D}=\{D_i,\mathbf{X}_i\}_{i=1}^n\).

\noindent For later use, define
\begin{equation*}
\begin{aligned}
    r_{qi} &\equiv   \exp(\beta_q^\top \mathbf{X}_i)\sum_{j=1}^{M} \Big[Y_{i}^\dagger (t_{(j)}) \,\rho_{q}\!\big(\boldsymbol{N_i^{\dagger}} (t_{(j)}^-);\boldsymbol{\alpha}_q\big)\, \Lambda_{0q} (\Delta t_{(j)})\Big],\\
    r_{0i} &\equiv  \exp(\beta_0^\top \mathbf{X}_i)\sum_{j=1}^{M} \Big[Y_{i}^\dagger (t_{(j)}) \,\rho_{0}\!\big(\boldsymbol{N_i^{\dagger}} (t_{(j)}^-);\boldsymbol{\gamma}\big)\, \Lambda_{00} (\Delta t_{(j)})\Big].
\end{aligned}
\end{equation*}

\noindent Then, the conditional posterior distributions of \(\Lambda_{0q}(\cdot)\) and \(\Lambda_{00}(\cdot)\) are as follows:
\begin{align}
    &p\!\left(\Lambda_{0q} (\Delta t_{(j)})\mid \Lambda_{00} (\Delta t_{(j)}), \mathbf{W}, \nu, \beta_q,\beta_0,\boldsymbol{\alpha}_q,\boldsymbol{\gamma}\right) \propto \mathrm{Gamma}\Bigg(\sum_{i=1}^{n}  N_{qi}^\dagger (\Delta t_{(j)}) + c_q \Lambda_{0q}^* (\Delta t_{(j)}), \notag\\
    &\hspace{6.0em} c_q + \sum_{i=1}^{n}  W_i \exp(\beta_q^\top \mathbf{X}_i) Y_{i}^\dagger (t_{(j)}) \,\rho_{q}\!\big(\boldsymbol{N_i^{\dagger}} (t_{(j)}^-);\boldsymbol{\alpha}_q\big)\Bigg),
\label{eq5}\\[2mm]
    &p\!\left(\Lambda_{00} (\Delta t_{(j)})\mid \Lambda_{0q} (\Delta t_{(j)}), \mathbf{W}, \nu, \beta_q,\beta_0,\boldsymbol{\alpha}_q,\boldsymbol{\gamma}\right) \propto \mathrm{Gamma}\Bigg(\sum_{i=1}^{n}  N_{0i}^\dagger (\Delta t_{(j)}) + c_0 \Lambda_{00}^* (\Delta t_{(j)}), \notag\\
    &\hspace{6.0em} c_0 + \sum_{i=1}^{n}  W_i \exp(\beta_0^\top \mathbf{X}_i) Y_{i}^\dagger (t_{(j)}) \,\rho_{0}\!\big(\boldsymbol{N_i^{\dagger}} (t_{(j)}^-);\boldsymbol{\gamma}\big)\Bigg).
\label{eq6}
\end{align}
Under an integrated squared-error loss function (posterior mean), we obtain
\begin{align}
    \widetilde{\Lambda}_{0q}(t\mid\mathbf{W},\Lambda_{00}, \nu, \beta_q,\beta_0,\boldsymbol{\alpha}_q,\boldsymbol{\gamma}) &= \sum_{j=1}^{M} \left[ \frac{\sum_{i=1}^{n} N_{qi}^\dagger (\Delta t_{(j)}) + c_q \Lambda_{0q}^* (\Delta t_{(j)})}{c_q + \sum_{i=1}^{n} W_i \exp(\beta_q^\top \mathbf{X}_i) Y_{i}^\dagger (t_{(j)}) \,\rho_{q}\!\big(\boldsymbol{N_i^{\dagger}} (t_{(j)}^-);\boldsymbol{\alpha}_q\big)} \right] \mathds{1}\{t_{(j)} \leq t\},
\label{eq7}\\[2mm]
    \widetilde{\Lambda}_{00}(t\mid\mathbf{W},\Lambda_{0q}, \nu, \beta_q,\beta_0,\boldsymbol{\alpha}_q,\boldsymbol{\gamma}) &= \sum_{j=1}^{M} \left[ \frac{\sum_{i=1}^{n} N_{0i}^\dagger (\Delta t_{(j)}) + c_0 \Lambda_{00}^* (\Delta t_{(j)})}{c_0 + \sum_{i=1}^{n} W_i \exp(\beta_0^\top \mathbf{X}_i) Y_{i}^\dagger (t_{(j)}) \,\rho_{0}\!\big(\boldsymbol{N_i^{\dagger}} (t_{(j)}^-);\boldsymbol{\gamma}\big)} \right] \mathds{1}\{t_{(j)} \leq t\}.
\label{eq8}
\end{align}
These are not yet estimators of \(\Lambda_{0q}(t)\) and \(\Lambda_{00}(t)\) since the \(W_i\) are unknown. However, we can obtain \(\hat{W}_i\) to replace \(W_i\) using the following conditional posterior distribution:
\begin{align}
    p\!\left(W_i \mid \Lambda_{0q} (\cdot), \Lambda_{00} (\cdot), \nu, \beta_q, \beta_0,\boldsymbol{\alpha}_q,\boldsymbol{\gamma}\right) &\propto W_i^{N_{. i}^\dagger (\cdot) + N_{0i}^\dagger (\cdot)} \exp\!\big(-W_i (r_{.i} + r_{0i})\big) \times g_{W_i}(w_i \mid \nu) \notag \\
    &\propto \mathrm{Gamma} \Big(N_{.i}^\dagger (\cdot) + N_{0i}^\dagger (\cdot) + \nu,\, r_{.i} + r_{0i} + \nu \Big),
\label{eq9}
\end{align}
where \(N_{. i}^\dagger (\cdot) = \sum_{q=1}^{Q} N_{qi}^\dagger (\cdot)\), \(r_{.i} = \sum_{q=1}^{Q} r_{qi}\), and \( g_{W_i}(w_i\mid\nu)  \propto W_i^{\nu - 1} \exp(-\nu W_i)\). The posterior mean is
 \begin{equation}
     \hat{W}_i = \frac{N_{.i}^\dagger (\cdot) + N_{0i}^\dagger (\cdot) +\nu}{r_{.i} + r_{0i} + \nu}, \quad i=1,2,\dots,n.
     \label{eq10}
 \end{equation}
Thus, we have closed-form Bayes plug-in estimators of \(\Lambda_{0q}(t)\) and \(\Lambda_{00}(t)\) given by
\begin{align}
    \hat{\Lambda}_{0q}(t\mid\Lambda_{00}, \nu, \beta_q,\beta_0,\boldsymbol{\alpha}_q,\boldsymbol{\gamma}) &= \sum_{j=1}^{M} \left[ \frac{\sum_{i=1}^{n} N_{qi}^\dagger (\Delta t_{(j)}) + c_q \Lambda_{0q}^* (\Delta t_{(j)})}{c_q + \sum_{i=1}^{n} \hat{W}_i \exp(\beta_q^\top \mathbf{X}_i) Y_{i}^\dagger (t_{(j)}) \,\rho_{q}\!\big(\boldsymbol{N_i^{\dagger}} (t_{(j)}^-);\boldsymbol{\alpha}_q\big)} \right] \mathds{1}\{t_{(j)} \leq t\},
\label{eq11}\\[2mm]
    \hat{\Lambda}_{00}(t\mid\Lambda_{0q}, \nu, \beta_q,\beta_0,\boldsymbol{\alpha}_q,\boldsymbol{\gamma}) &= \sum_{j=1}^{M} \left[ \frac{\sum_{i=1}^{n} N_{0i}^\dagger (\Delta t_{(j)}) + c_0 \Lambda_{00}^* (\Delta t_{(j)})}{c_0 + \sum_{i=1}^{n} \hat{W}_i \exp(\beta_0^\top \mathbf{X}_i) Y_{i}^\dagger (t_{(j)}) \,\rho_{0}\!\big(\boldsymbol{N_i^{\dagger}} (t_{(j)}^-);\boldsymbol{\gamma}\big)} \right] \mathds{1}\{t_{(j)} \leq t\}.
\label{eq12}
\end{align}
We can recover the Breslow–Aalen–type estimator of the baseline cumulative hazard function from Eqs.~(\ref{eq11}) and (\ref{eq12}) by letting the precision parameters \( c_q \to 0 \) and \( c_0 \to 0 \). The conditional posterior distribution of the frailty parameter \( \nu \) is given by
\begin{align}
    p\!\left(\nu \mid \Lambda_{0q} (\cdot), \Lambda_{00} (\cdot), \beta_0, \beta_q, \boldsymbol{\alpha}_q,\boldsymbol{\gamma}\right) &\propto L_m \!\left(\Lambda_{0q} (\cdot), \Lambda_{00} (\cdot), \nu, \beta_q, \beta_0\right) \times \pi(\nu) \notag \\
    &\propto \prod_{i=1}^{n} \Bigg[ \frac{\Gamma \!\big( N_{.i}^\dagger (\cdot) + N_{0i}^\dagger (\cdot) + \nu \big)}{\big(r_{.i} + r_{0i} + \nu\big)^{\,N_{.i}^\dagger (\cdot) + N_{0i}^\dagger (\cdot) + \nu}} \times \frac{\nu^\nu}{\Gamma(\nu)} \Bigg] \times \pi(\nu),
\label{eq13}
\end{align}
where \(L_m\) denotes the marginal likelihood and \(\pi(\nu) \propto \nu^{\zeta-1} \exp(-\eta \nu)\).

\noindent The conditional posterior distributions of the regression parameters \(\beta_q\) and \(\beta_0\) are given by
\begin{align}
    p\!\left(\beta_q \mid \Lambda_{0q} (\cdot), \Lambda_{00} (\cdot), \mathbf{W}, \nu, \beta_0,\boldsymbol{\alpha}_q,\boldsymbol{\gamma}\right) &\propto L_m \!\left(\beta_q, \Lambda_{0q} (\cdot), \Lambda_{00} (\cdot), \nu, \beta_0\right) \times \pi(\beta_q) \notag \\
    &\propto \exp \Bigg[ \sum_{i=1}^{n} N_{qi}^\dagger (\cdot) \cdot (\beta_q^\top \mathbf{X}_i) - \sum_{i=1}^{n} W_i \cdot r_{qi} \Bigg] \times \pi(\beta_q),
\label{eq14}
\end{align}
\begin{align}
    p\!\left(\beta_0 \mid \Lambda_{0q} (\cdot), \Lambda_{00} (\cdot), \mathbf{W}, \nu,  \beta_q,\boldsymbol{\alpha}_q,\boldsymbol{\gamma}\right) &\propto L_m \!\left(\beta_0, \Lambda_{0q} (\cdot), \Lambda_{00} (\cdot), \nu, \beta_q\right) \times \pi(\beta_0) \notag \\
    &\propto \exp \Bigg[ \sum_{i=1}^{n} N_{0i}^\dagger (\cdot) \cdot (\beta_0^\top \mathbf{X}_i) - \sum_{i=1}^{n} W_i \cdot r_{0i} \Bigg] \times \pi(\beta_0),
\label{eq15}
\end{align}
where \(\pi(\beta_q) \propto \exp\big[-\tfrac{1}{2} (\beta_q-\mu_{\beta_q})^\top\Sigma_{\beta_q}^{-1} (\beta_q-\mu_{\beta_q})\big]\) and \(\pi(\beta_0) \propto \exp\big[-\tfrac{1}{2} (\beta_0-\mu_{\beta_0})^\top\Sigma_{\beta_0}^{-1} (\beta_0-\mu_{\beta_0})\big]\).

\noindent For the dynamic parameters, the conditional posterior kernels (non-conjugate) are
\begin{align}
&p(\boldsymbol{\alpha}_q \mid \Lambda_{0q},\Lambda_{00},\mathbf{W},\beta_q,\beta_0,\nu,\boldsymbol{\gamma})
\;\propto\;
\exp\Bigg\{
  \sum_{i=1}^{n}\sum_{j=1}^{M}
    N^\dagger_{qi}(\Delta t_{(j)})\,
    \log\!\Big(\rho_{q}\!\big(\boldsymbol{N_i^{\dagger}} (t_{(j)}^-);\boldsymbol{\alpha}_q\big)\,\Lambda_{0q}(\Delta t_{(j)})\Big) \nonumber\\
&\hspace{10.3em}
  -\,\sum_{i=1}^{n}W_i\,\exp(\mathbf{X}_i^\top\beta_q)
    \sum_{j=1}^{M}
      Y^\dagger_i(t_{(j)})\,
      \rho_{q}\!\big(\boldsymbol{N_i^{\dagger}} (t_{(j)}^-);\boldsymbol{\alpha}_q\big)\,
      \Lambda_{0q}(\Delta t_{(j)})
\Bigg\}\times \pi(\boldsymbol{\alpha}_q),
\label{eq16}
\end{align}
\begin{align}
&p(\boldsymbol{\gamma} \mid \Lambda_{0q},\Lambda_{00},\mathbf{W},\beta_q,\beta_0,\nu,\boldsymbol{\alpha}_q)
\;\propto\;
\exp\Bigg\{
  \sum_{i=1}^{n}\sum_{j=1}^{M}
    N^\dagger_{0i}(\Delta t_{(j)})\,
    \log\!\Big(\rho_{0}\!\big(\boldsymbol{N_i^{\dagger}} (t_{(j)}^-);\boldsymbol{\gamma}\big)\,\Lambda_{00}(\Delta t_{(j)})\Big) \nonumber \\
&\hspace{10.3em}
  -\,\sum_{i=1}^{n}W_i\,\exp(\mathbf{X}_i^\top\beta_0)
    \sum_{j=1}^{M}
      Y^\dagger_i(t_{(j)})\,
      \rho_{0}\!\big(\boldsymbol{N_i^{\dagger}} (t_{(j)}^-);\boldsymbol{\gamma}\big)\,
      \Lambda_{00}(\Delta t_{(j)})
\Bigg\}\times \pi(\boldsymbol{\gamma}),
\label{eq17}
\end{align}
\noindent where
\begin{equation}
\begin{aligned}
\pi(\boldsymbol{\alpha}_q)
&\propto
\exp\!\Big[
(a-1)\sum_{\ell=1}^{Q}\log(\alpha_{q\ell})
- b\sum_{\ell=1}^{Q}\alpha_{q\ell}
\Big],\\
\pi(\boldsymbol{\gamma})
&\propto
\exp\!\Big[
(c-1)\sum_{\ell=1}^{Q}\log(\gamma_{\ell})
- d\sum_{\ell=1}^{Q}\gamma_{\ell}
\Big].
\end{aligned}
\label{eq18}
\end{equation}

\subsection{MCMC Algorithm}
The conditional posteriors of \(\Lambda_{0q}\), \(\Lambda_{00}\), and \(W_i\) in Eqs.~(\ref{eq5}), (\ref{eq6}), and (\ref{eq9}) are available in closed form and can be sampled directly or updated via the Bayes estimators in Eqs.~(\ref{eq11}), (\ref{eq12}), and (\ref{eq10}). The frailty variance parameter \(\nu\) is updated using a Metropolis–Hastings (MH) step targeting Eq.~(\ref{eq13}). In contrast, the regression parameters \(\beta_q,\beta_0\) and the dynamic past-event parameters \(\boldsymbol{\alpha}_q,\boldsymbol{\gamma}\) are updated using Differential Evolution MCMC (DEMCMC), a population-based proposal mechanism, targeting Eqs.~(\ref{eq14}), (\ref{eq15}), (\ref{eq16}), and (\ref{eq17}). After the first introduction, we refer to this scheme simply as DEMCMC.

\noindent\textbf{Pseudocode.}
\begin{enumerate}
\renewcommand\labelenumi{(\roman{enumi})}
\item \textbf{Initialization:} Set \(\Lambda_{0q}^{(0)}(\cdot), \Lambda_{00}^{(0)}(\cdot), W^{(0)}, \nu^{(0)}, 
      \beta_q^{(0)}, \beta_0^{(0)}, \boldsymbol{\alpha}_q^{(0)}, \boldsymbol{\gamma}^{(0)}\).
\item For \(b=1,\dots,B\) iterations do:
  \begin{enumerate}
  \renewcommand\labelenumii{(\alph{enumii})}
    \item For each \(q\), update \(\Lambda_{0q}^{(b)}\) either by drawing samples using Eq.~(\ref{eq5}) or by updating \(\widetilde{\Lambda}_{0q}\) via Eq.~(\ref{eq7}) with current \(W^{(b-1)}\), \(\beta_q^{(b-1)}\), and \(\boldsymbol{\alpha}_q^{(b-1)}\).
    \item Update \(\Lambda_{00}^{(b)}\) analogously using Eq.~(\ref{eq6}) or Eq.~(\ref{eq8}) with current \(W^{(b-1)}\), \(\beta_0^{(b-1)}\), and \(\boldsymbol{\gamma}^{(b-1)}\).
    \item Compute \(r_{qi}\) and \(r_{0i}\) using the current \(\Lambda_{0q}^{(b)}\), \(\Lambda_{00}^{(b)}\), and 
          \(\beta_q^{(b-1)}, \beta_0^{(b-1)}\), together with the dynamic factors 
          \(\rho_q(\cdot;\boldsymbol{\alpha}_q^{(b-1)})\) and \(\rho_0(\cdot;\boldsymbol{\gamma}^{(b-1)})\)
          as defined preceding Eq.~(\ref{eq5}).
    \item Sample \(\nu^{(b)}\) using an MH update targeting Eq.~(\ref{eq13}).
    \item Update \(W_i^{(b)}\) for \(i=1,\dots,n\) either by drawing from Eq.~(\ref{eq9}) or by using the posterior mean in Eq.~(\ref{eq10}).
    \item Update \(\beta_q^{(b)}\) using DEMCMC targeting Eq.~(\ref{eq14}).
    \item Update \(\beta_0^{(b)}\) using DEMCMC targeting Eq.~(\ref{eq15}).
    \item For each \(q\), update \(\boldsymbol{\alpha}_q^{(b)}\) using DEMCMC targeting Eq.~(\ref{eq16}) with prior in Eq.~(\ref{eq18}).
    \item Update \(\boldsymbol{\gamma}^{(b)}\) using DEMCMC targeting Eq.~(\ref{eq17}) with prior in Eq.~(\ref{eq18}).
  \end{enumerate}
\item After burn-in and thinning, compute \(\hat{\Lambda}_{0q}\) and \(\hat{\Lambda}_{00}\) using 
      Eqs.~(\ref{eq11})--(\ref{eq12}), and evaluate any derived quantities (e.g., baseline survival). 
      Summarize posterior draws for point estimates, credible intervals, and diagnostics.
\end{enumerate}

\section{Simulation Study}
\label{sec:sim}
To evaluate the finite-sample performance of the proposed Bayesian dynamic frailty model, we conducted a comprehensive simulation study under realistic settings. For computational tractability, we considered a special case of the general dynamic model in which the effect of the past event was restricted to influence only future events of the same type, rather than allowing cross-type dependencies. Under this special-case formulation, the recurrent-event history effects are summarized by the three-dimensional vector $\boldsymbol{\alpha} = (\alpha_{1}, \alpha_{2}, \alpha_{3})$, where $\alpha_{q}$ governs how prior events of type-$q$ alter the future hazard of type-$q$. Similarly, the dynamic effects linking past recurrent events to the terminal event are represented by $\boldsymbol{\gamma} = (\gamma_{1}, \gamma_{2}, \gamma_{3})$, where $\gamma_{q}$ quantifies how past type-$q$ events modify the terminal event hazard.  This setup simplifies the dependence structure while preserving the essential past event  dynamic effects on future subsequent of same type as well as association between recurrent events and the terminal event.

\subsection{Simulation Design}

We generated independent datasets of size \(n \in \{100,200\}\). Each subject contributed two independent covariates: a binary variable \(X_1 \sim \mathrm{Bernoulli}(0.5)\) and a continuous variable \(X_2 \sim \mathcal{N}(0,1)\). Subject-specific heterogeneity was introduced through a shared frailty term \(W_i \sim \mathrm{Gamma}(\nu,\nu)\) with \(\nu \in \{2,4\}\); smaller values of \(\nu\) correspond to stronger dependence among the recurrent event processes and the terminal event.

Baseline cumulative hazard functions for the three recurrent event types, \(\Lambda_{01}(t)\), \(\Lambda_{02}(t)\), and \(\Lambda_{03}(t)\), were generated from Weibull distributions with a common shape parameter \(\gamma^\ast \in \{0.9,1.1\}\). A value of \(\gamma^\ast < 1\) produces a decreasing failure rate over time, where the hazard diminishes as time progresses, representing a decreasing failure rate process (DFR). In contrast, \(\gamma^\ast > 1\) yields an increasing failure rate, where hazards rise over time, representing an increasing failure rate process (IFR). The corresponding scale parameters for the three recurrent event types were set to \(1.2\), \(1.3\), and \(1.4\). The terminal event baseline hazard \(\Lambda_{00}(t)\) also followed a Weibull distribution with the same shape parameter \(\gamma^\ast\) and a scale parameter of \(2.2\). All subjects were administratively followed until \(\tau = 3\), and independent censoring times were drawn from a \(\mathrm{Uniform}(1,3)\) distribution.

To allow recurrence history to influence subsequent event risk, we incorporated dynamic past-event effects. For the recurrent event types, we specified \(\boldsymbol{\alpha} = (0.35, 0.30, 0.25)\), representing increasing contributions of past events to future hazards. The terminal event model included analogous dynamic effects through \(\boldsymbol{\gamma} = (0.20, 0.15, 0.10)\), permitting the terminal hazard to increase with accumulated recurrence activity. Covariate effects followed proportional hazards structures with regression coefficients \((-0.40,0.35)\), \((-0.30,0.25)\), \((-0.20,0.15)\), and \((-0.10,0.10)\) assigned to the three recurrent types and the terminal event.

Gamma–process priors with precision \(c = 0.1\) were assigned to all baseline hazard increments. Regression coefficients were given independent \(N(0,1)\) priors, while the dynamic parameters \(\alpha_q\) and \(\gamma_q\) each followed independent \(\mathrm{Gamma}(0.5,2)\) priors. The frailty parameter \(\nu\) received a \(\mathrm{Gamma}(1,1)\) prior.

To assess robustness, we reanalyzed each setting under deliberately misspecified priors. For the recurrent event baselines, \(\Lambda_{01}(t)\), \(\Lambda_{02}(t)\), and \(\Lambda_{03}(t)\) were assigned \(\mathrm{Exponential}(10)\), \(\mathrm{Exponential}(11)\), and \(\mathrm{Exponential}(12)\) priors, and the terminal baseline \(\Lambda_{00}(t)\) was given an \(\mathrm{Exponential}(13)\) prior. In all four baselines, the precision parameters were set to \(c_q = c_0 = 0.01\). We further performed a sensitivity analysis by varying the prior variances for the regression and frailty parameters. The “standard” prior used \((\sigma^2_\beta = 1, \sigma^2_\nu = 0.5)\) in our analysis and was compared with strong \((0.25,0.25)\), weak \((2.25,1)\), and vague \((9,2)\) alternatives.

For each combination of \((n,\nu,\gamma^\ast)\), we generated 500 replicated datasets and analyzed them using an MCMC procedure with 5{,}000 iterations, a burn-in of 2{,}000, and thinning of 5 (Tables~\ref{tab:gamma1.1}-\ref{tab:gamma0.9}, Figures~\ref{fig:Figure_2_dm.pdf}-\ref{fig:Figure_3_dm.pdf}). Convergence was monitored using the Gelman--Rubin statistic and effective sample size diagnostics (Table~\ref{tab:table3}). Performance was evaluated through empirical bias, standard deviation, RMSE, and 95\% coverage probability for the frailty parameter \(\nu\), the regression coefficients \(\beta_q,\beta_0\), and the dynamic parameters \((\boldsymbol{\alpha},\boldsymbol{\gamma})\). To assess recovery of the baseline hazard functions, we also computed pointwise RMSE for the estimated marginal survival functions.

\subsection{Simulation Results}

We evaluated the characteristics of the proposed Bayesian dynamic frailty model across combinations of the frailty parameter \((\nu \in \{2,4\})\), sample size \((n \in \{100,200\})\), and baseline hazard shape parameter \((\gamma^\ast \in \{1.1,0.9\})\), corresponding to IFR and DFR processes. For each setting, we assessed estimation accuracy for the regression parameters \((\beta_{11},\beta_{12},\beta_{21},\beta_{22},\beta_{31},\beta_{32},\beta_{01},\beta_{02})\), the dynamic parameters for the recurrent processes \((\alpha_1,\alpha_2,\alpha_3)\), the terminal dynamic parameters \((\gamma_1,\gamma_2,\gamma_3)\), and the frailty parameter \(\nu\). We also computed pointwise RMSE for the estimated marginal survival functions of all recurrent event types and the terminal event. Results for the IFR and DFR scenarios are summarized in Tables~\ref{tab:gamma1.1} and \ref{tab:gamma0.9}, respectively.

Across both failure-rate structures, the estimation of the frailty parameter \(\nu\) shows the strongest sensitivity to dependence strength and sample size. Under IFR, the estimator exhibits substantial negative bias when \(\nu = 2\), particularly with \(n=100\), where the bias reaches \(-0.836\) (RMSE = 0.911); increasing the sample size to \(n=200\) reduces but does not eliminate this difficulty (RMSE = 1.259) (Table~\ref{tab:gamma1.1}). When \(\nu = 4\), estimation accuracy improves substantially, with RMSE values between 0.464 and 0.650 and coverage probabilities close to 0.99 (Table~\ref{tab:gamma1.1}). Similar patterns appear in the DFR setting, where \(\nu = 2\) remains challenging while \(\nu = 4\) yields more stable and precise estimates (Table~\ref{tab:gamma0.9}). These findings highlight the difficulty of recovering frailty effects under strong dependence and smaller sample sizes.

Estimation of the regression coefficients is excellent across all scenarios. Under IFR with \(n=100\), biases for the type~1 and type~2 regression coefficients lie between \(-0.056\) and \(0.031\), with similarly small values for the type~3 and terminal-event coefficients, and these biases and RMSE values decrease further when \(n=200\) (Table~\ref{tab:gamma1.1}). The DFR results closely mirror the IFR behavior, with small bias and modest RMSE for all regression parameters across \((n,\nu)\) combinations (Table~\ref{tab:gamma0.9}). These patterns demonstrate that the proposed Bayesian method provides stable and precise estimation of covariate effects in the multitype recurrent event framework.

The recurrent-process dynamic parameters \((\alpha_1,\alpha_2,\alpha_3)\) show the strongest and most stable performance among all estimators. Across IFR and DFR settings, biases remain close to zero, RMSE values range from 0.067 to 0.121, and coverage probabilities are near or above 0.98 (Tables~\ref{tab:gamma1.1}–\ref{tab:gamma0.9}). For example, under IFR with \(n=200\) and \(\nu=4\), the RMSE values for \((\alpha_1,\alpha_2,\alpha_3)\) are 0.087, 0.083, and 0.067 (Table~\ref{tab:gamma1.1}). These results indicate that within-type dynamic past recurrent-event feedback effects are recovered with high precision.

The dynamic terminal-event parameters \((\gamma_1,\gamma_2,\gamma_3)\) are also estimated accurately. Under IFR with \(n=100\), RMSE ranges from 0.135 to 0.180, decreasing to 0.089–0.127 when \(n=200\), with coverage probabilities between 0.93 and 0.99 (Table~\ref{tab:gamma1.1}). The DFR results show the same improvement with increasing sample size and weaker dependence, with similarly high coverage across all combinations of \((n,\nu)\) (Table~\ref{tab:gamma0.9}). These findings confirm reliable posterior recovery of the cross-process dynamic effects governing the terminal event.

Overall, increasing sample size and using higher frailty values (\(\nu=4\)) consistently improve estimation precision across all parameter groups. Larger \(n\) leads to reduced SD and RMSE for all parameters, and higher \(\nu\) improves estimation relative to \(\nu=2\) in both IFR and DFR settings (Tables~\ref{tab:gamma1.1}–\ref{tab:gamma0.9}). Coverage probabilities for the regression and dynamic parameters remain close to the nominal 95\% level, with only mild undercoverage observed for the frailty parameter when \(n=100\) and \(\nu=2\). Comparable conclusions hold under the DFR setting (Table~\ref{tab:gamma0.9}).

\begin{figure}[H]
    \centering
    \vspace{-0.2cm}\includegraphics[width=1\textwidth,height=0.9\textheight]{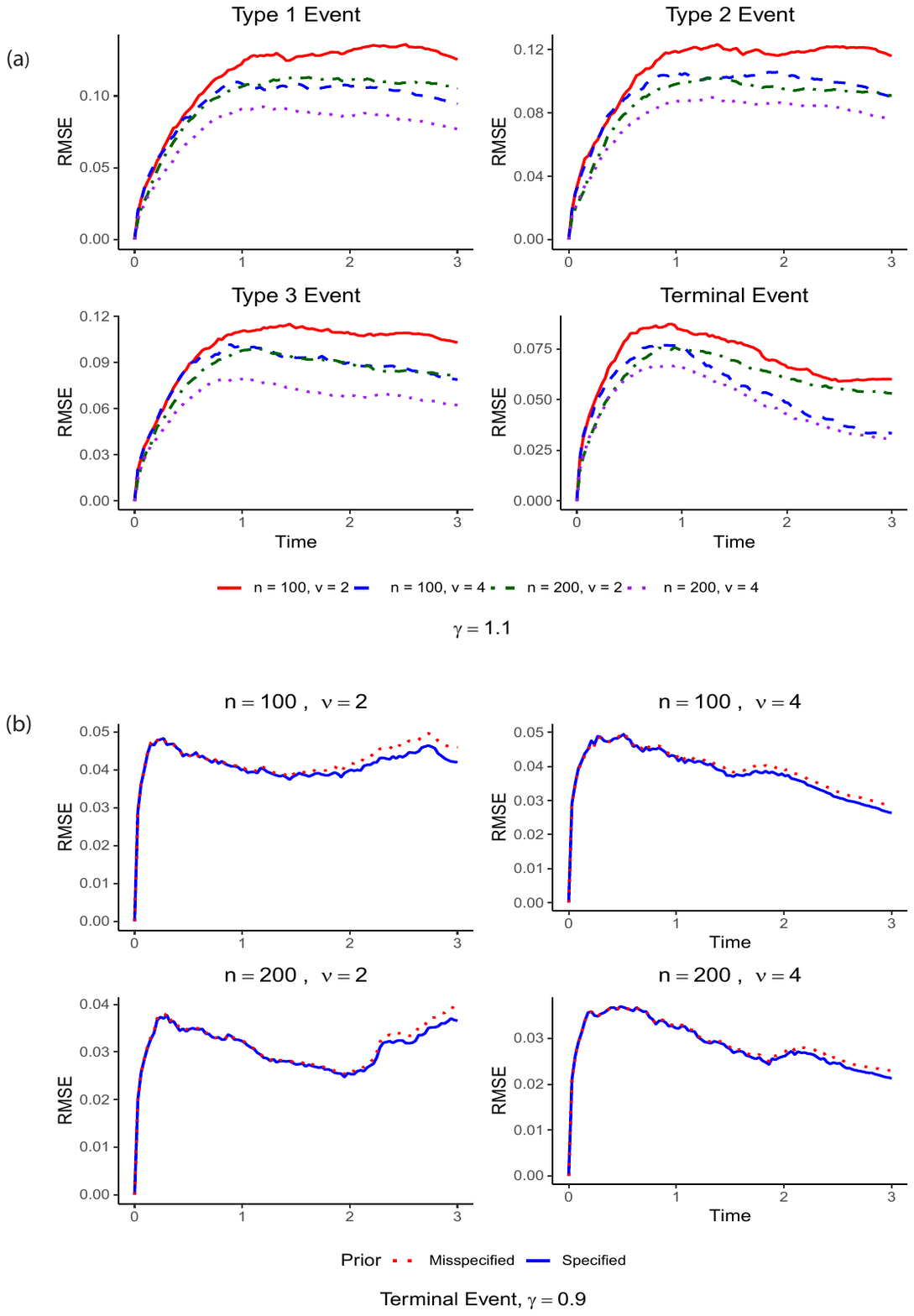}
    \caption{(a) RMSE of the estimated baseline survival functions for the three recurrent events and the terminal event under increasing failure rate ($\gamma^\ast = 1.1$), evaluated across sample sizes $n \in \{100,200\}$ and frailty levels $\nu \in \{2,4\}$. (b) RMSE of the terminal event estimated baseline survival function under correctly specified and misspecified priors when $\gamma = 0.9$, demonstrating robustness to prior misspecification.}
    \label{fig:Figure_2_dm.pdf}
\end{figure}

Figure~\ref{fig:Figure_2_dm.pdf}(a) presents pointwise RMSE for the estimated baseline survival functions under IFR \((\gamma^\ast = 1.1)\). Across all event types, the largest RMSE values occur when \(n=100\) and \(\nu=2\), whereas the smallest occur for \(n=200\) and \(\nu=4\), illustrating the expected gains in baseline estimation accuracy with increased sample size and weaker dependence. Figure~\ref{fig:Figure_2_dm.pdf}(b) depicts the robustness evaluation under baseline prior misspecification: even when exponential working means are used instead of the true Weibull structure, the RMSE curves for the terminal event under DFR \((\gamma^\ast = 0.9)\) remain nearly identical across all \((n,\nu)\) combinations. These results indicate that the Bayesian model is highly robust to moderate baseline prior misspecification.

\begin{figure}[H]
    \centering
    \vspace{-0.2cm}\includegraphics[width=0.9\textwidth,height=0.75\textheight]{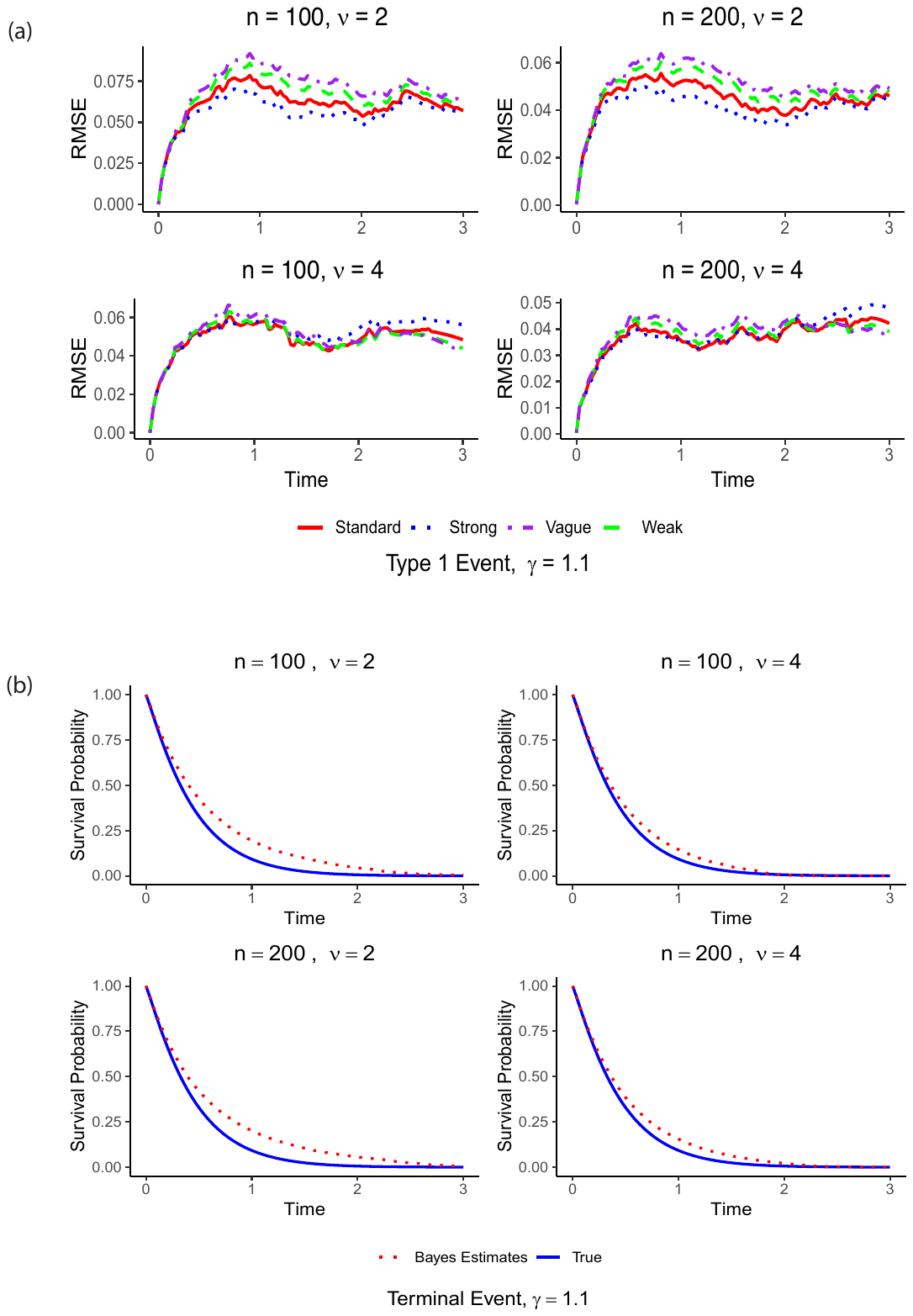}
    \caption{(a) Comparison of RMSE for the Type~1 event baseline survival estimation under increasing failure rate ($\gamma^\ast=1.1$) for standard, strong, vague, and weak priors across $(n,\nu)$ combinations showing the robustness to prior choice. (b) Corresponding terminal-event baseline survival Bayes estimates against the true function, demonstrating overall accuracy}
    \label{fig:Figure_3_dm.pdf}
\end{figure}

Figure~\ref{fig:Figure_3_dm.pdf}(a) summarizes the prior sensitivity analysis for type~1 recurrent events under IFR. The RMSE curves corresponding to the Standard, Strong, Weak, and Vague priors lie closely together over time and across all sample sizes and frailty strengths, with only minor deviations at early time points. This suggests that inference is only weakly sensitive to the prior variance choices. Figure~\ref{fig:Figure_3_dm.pdf}(b) compares the estimated and true survival functions for the terminal event under IFR \((\gamma^\ast = 1.1)\). The Bayesian estimates closely track the true survival curve for all \((n,\nu)\) combinations, with the largest discrepancies observed when \(n=100\) and \(\nu=2\); these gaps narrow substantially for larger sample sizes and weaker dependence. Together, these findings demonstrate the strong predictive accuracy of the proposed dynamic frailty model.

Convergence diagnostics are reported in Table~\ref{tab:table3}. All parameters exhibit Gelman--Rubin \(\hat{R}\) values at or near 1.0, indicating strong mixing of the MCMC chains, and effective sample sizes typically exceed 10\% of the total post–burn-in draws. These diagnostics confirm that the Bayesian parameter estimates are stable, robust, and well supported by the MCMC sampling process.

All analyses were conducted in R version~4.4.2. Custom R functions were written to implement the proposed Bayesian dynamic frailty model using MCMC.

\section{Application}
\label{sec:app}
ALLHAT Collaborative Research Group \citep{ALLHAT2002} report the Antihypertensive and Lipid-Lowering Treatment to Prevent Heart Attack Trial (ALLHAT), a large-scale, double-blind, randomized trial conducted at 623 centers in North America from February 1994 to March 2002. The trial, funded by the National Heart, Lung, and Blood Institute (NHLBI), compared the effectiveness of antihypertensive medications for the prevention of major coronary events in high-risk patients. The study enrolled 33{,}357 hypertensive participants with at least one additional cardiovascular risk factor: 15{,}255 (45.73\%) received chlorthalidone, 9{,}048 (27.12\%) received amlodipine, and 9{,}054 (27.15\%) received lisinopril. The average follow-up was 4.9 years \citep{wright2005outcomes}, and 95\% of participants completed follow-up assessments.

During follow-up, five major cardiovascular events were tracked: myocardial infarction, stroke, congestive heart failure, angina, and peripheral arterial disease. Overall, 4.0\% experienced myocardial infarction, 3.4\% stroke, 4.4\% congestive heart failure, 7.1\% angina, and 1.6\% peripheral arterial disease. These outcomes have distinct clinical courses and management pathways, so distinguishing them is important for risk evaluation and treatment comparisons.

Myocardial infarction and stroke typically arise abruptly from acute thrombotic or ischemic events and require immediate intervention (e.g., thrombolysis, percutaneous coronary intervention (PCI), time-sensitive stroke management) \citep{benjamin2019heart,fuster2014acute}. In contrast, chronic cardiovascular dysfunction—such as congestive heart failure, angina, and peripheral arterial disease—reflects sustained vascular injury, impaired flow regulation, and endothelial dysfunction, and is managed with lifestyle modification, medical therapy (e.g., beta-blockers, angiotensin-converting enzyme (ACE) inhibitors, statins), and, in selected cases, revascularization \citep{Roffi2016,GerhardHerman2017,mh20172017,fowkes2013comparison}. Guided by these patterns, we classify outcomes into acute cardiovascular events (ACEs; myocardial infarction and stroke) and chronic cardiovascular events (CCEs; congestive heart failure, angina, and peripheral arterial disease), which clarifies immediate threats versus long-term burden and supports stratified risk modeling \citep{Go2013,Townsend2016}.

Among the 33{,}357 participants, 1{,}423 (4.26\%) experienced acute cardiovascular events (ACE), 2{,}890 (8.48\%) experienced chronic cardiovascular events (CCE), and 931 (2.79\%) experienced both during follow-up. In addition, 4{,}929 (14.8\%) died from all causes, underscoring the burden of cardiovascular disease among hypertensive patients. Multiple cardiovascular incidents may accelerate progression toward higher mortality, suggesting interdependence between recurrent and terminal events and motivating a multitype recurrent-event framework with frailty. Note that primary analysis of this clinical trial considered only first event of the recurrent of same type  and analyzed each type of outcome separately \citep{ALLHAT2002}. In this paper we considered all repeated events of the same type as well as multitype events and terminal event jointly.

We fit the semiparametric joint dynamic model using MCMC algorithm. Regression coefficients used independent normal priors $\beta_{ql}\sim\mathcal{N}(0,10)$ (mean $0$, variance $10$) 
for all covariates except race, for which informative priors were specified: $\beta_{13},\beta_{23},\beta_{03}\sim\mathcal{N}(0.2,0.005)$. The dynamic parameters were assigned independent gamma priors 
$\alpha_{q},\gamma_{q}\sim\mathrm{Gamma}(0.1,0.1)$ (mean $1$, variance $10$) where \(q\) are ACE and CCE. The frailty parameter had a gamma prior 
$\nu\sim\mathrm{Gamma}(0.1,0.1)$. At each iteration, all 
regression coefficients and dynamic parameters were updated using Differential Evolution MCMC (DEMCMC), whereas $\nu$ was updated via a Metropolis--Hastings step. We ran 5{,}000 iterations, discarded the first 2{,}000 as burn-in, and retained every third draw, yielding 1{,}000 posterior samples.

In Table~\ref{tab:table4}, both amlodipine and lisinopril (reference = chlorthalidone) are associated with higher risk of experiencing an ACE event, with hazard ratios of 1.92 (95\% CI: 1.61--2.10) and 2.45 (95\% CI: 2.24--2.68), respectively. Black participants
show substantially elevated ACE risk (HR = 4.20; 95\% CI: 3.75--4.62), and older age is also linked to higher ACE hazard (HR = 4.65; 95\% CI: 4.33--4.95). For CCE events, amlodipine and lisinopril again demonstrate increased hazards relative to chlorthalidone, with hazard ratios of 1.82 (95\% CI: 1.69--1.97) and 2.11 (95\% CI: 1.95--2.28). Black race is associated with a markedly higher CCE hazard (HR = 4.59; 95\% CI: 4.16--5.04), and age shows a similar pattern 
(HR = 4.10; 95\% CI: 3.85--4.37). For the terminal event (Death), both amlodipine and lisinopril indicate increased mortality risk, with hazard ratios of 2.57 (95\% CI: 2.37--2.77) and 3.13 (95\% CI: 2.90--3.40). The effect of race is again pronounced (HR = 7.14; 95\% CI: 6.45--7.82), and age is strongly associated with death (HR = 7.20; 95\% CI: 6.74--7.67).

The frailty variance estimate 
$\nu=0.12$ (SE = 0.003) suggests considerable unobserved heterogeneity. The dynamic parameters describe how previous cardiovascular events alter the intensity of future events. Each prior acute cardiovascular event increases the intensity of experiencing another ACE by nearly 90\%, indicating moderate within-process reinforcement. The effect is even stronger for chronic events: each prior CCE raises the intensity of another future CCE by approximately 180\%, suggesting substantial persistence in chronic disease progression. In contrast, the dynamic contributions to mortality are more modest. Each previous ACE increases the death intensity by about 12\%, while the 
effect of prior CCEs on mortality is negligible (less than a 1\% increase per event). Together, these findings show that recurrent cardiovascular events—particularly chronic 
events—strongly reinforce their own future risk but exert relatively limited direct influence on mortality once baseline covariates and frailty are considered.

\begin{figure}[H]
    \centering
    \vspace{-0.2cm}\includegraphics[width=0.75\textwidth,height=0.90\textheight]{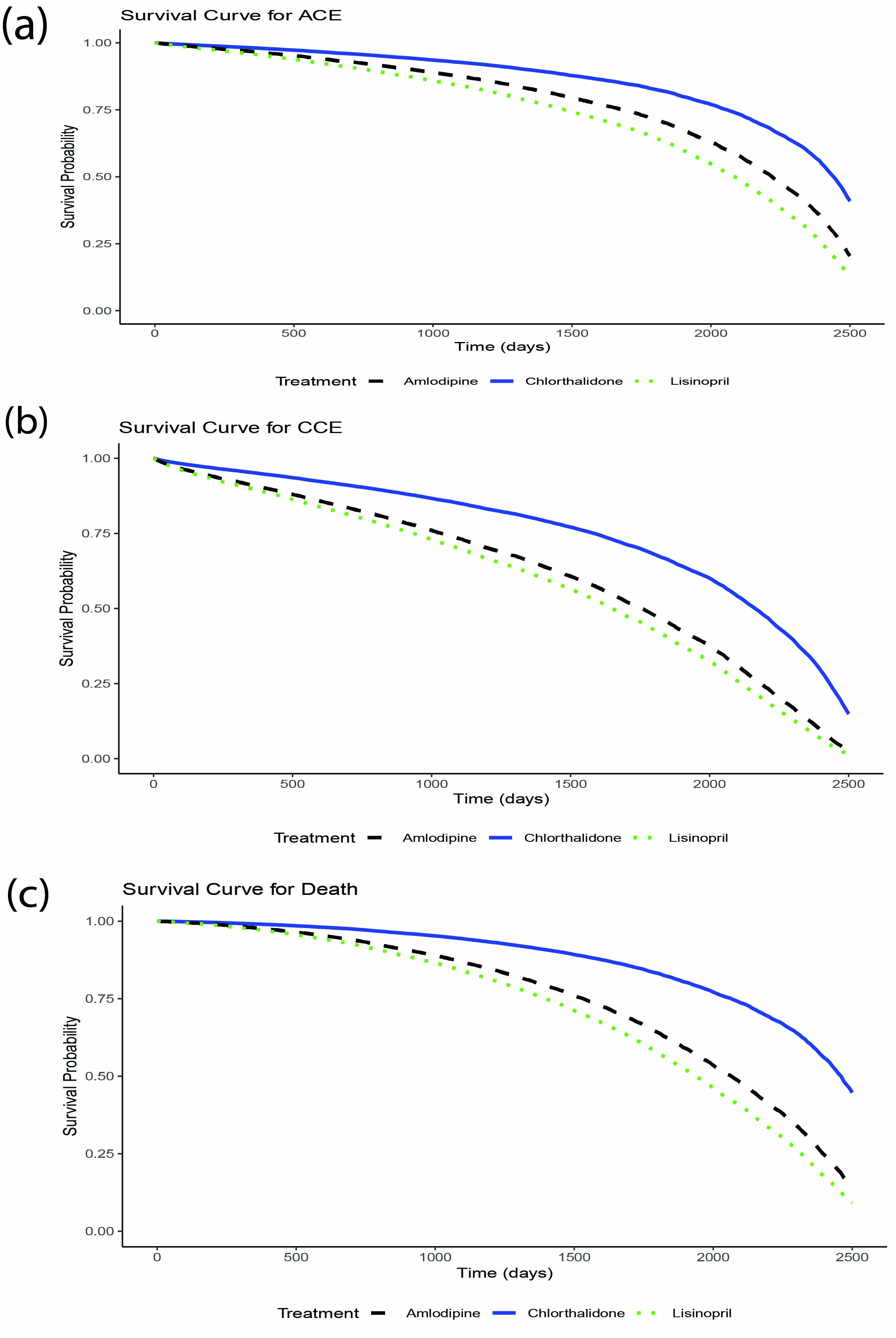}
    \caption{Treatment-specific survival probabilities for (a) acute cardiovascular events, (b) chronic cardiovascular events, and (c) death, comparing Chlorthalidone, Amlodipine, and Lisinopril.}
    \label{fig:Figure_4.jpg}
\end{figure}

Based on treatment groups, Figure~\ref{fig:Figure_4.jpg}(a--c) displays survival probabilities for ACE, CCE, and death. For ACE (panel a), chlorthalidone (reference) yields the highest survival; amlodipine and lisinopril exhibit nearly overlapping, lower survival. A similar ordering appears for CCE (panel b). For death (panel c), chlorthalidone again shows the highest survival, followed by amlodipine, with lisinopril lowest. Overall, chlorthalidone is more effective than amlodipine followed by lisinopril in reducing  ACE and CCE events and lowering the risks of mortality.

\section{Discussion}
\label{sec:conc}

We developed a semiparametric Bayesian joint dynamic model for multiple recurrent events and a terminal event. The framework combines a shared frailty to capture subject-level heterogeneity with simple, interpretable dynamic functions that allow past events to modify future risks. Flexible Gamma–process priors yield smooth baseline cumulative hazards with closed-form updating, providing stable computation even when event patterns are sparse or irregular. Across all simulation scenarios, the Bayesian estimator achieved low RMSE and near-nominal coverage for the regression and frailty parameters, produced accurate baseline hazard estimates, and remained stable under misspecification of the baseline hazard and under a wide range of prior choices. Convergence diagnostics consistently indicated well-mixed posterior samples and reliable inference.

The model was applied to the ALLHAT study to jointly analyze acute (ACE) and chronic (CCE) cardiovascular events along with all-cause mortality. The shared frailty captured substantial unobserved heterogeneity, and the dynamic component indicated that prior cardiovascular events meaningfully increase the risk of subsequent events. Treatment comparisons were consistent across outcomes: amlodipine and lisinopril were associated with higher risks of ACE, CCE, and mortality, whereas chlorthalidone demonstrated a protective profile across all three event types. Age and race (Black vs.\ White) showed strong and directionally consistent associations, with especially pronounced effects on mortality, aligning with previously reported findings from ALLHAT \citep{ALLHAT2002,wright2005outcomes}.

The race coefficient required careful treatment. Under weak priors, the estimates were unstable—a pattern expected when covariate overlap is limited and monotone-likelihood behavior arises in semiparametric hazard models \citep{HeinzeSchemper2001,GreenlandMansourniaAltman2016,RipattiPalmgren2000}. Diagnostics supported this interpretation: Black participants had much higher mean linear predictors than White participants for ACE (1.73 vs.\ 0.49), CCE (1.77 vs.\ 0.43), and death (2.37 vs.\ 0.65), with standardized mean differences of 0.78–0.93, indicating substantial separation in risk. To address this, we stabilized the race effect using an informative normal prior calibrated through prior–predictive checks to yield plausible hazard ratios \citep{Firth1993,GelmanJakulinPittauSu2008}. Sensitivity analyses using wider normal and Student-\(t\) priors showed that estimates for treatment, age, and frailty were robust; only the magnitude of the race effect varied with the level of regularization, consistent with established results \citep{Firth1993,HeinzeSchemper2001}.

Several limitations merit mention. The model assumes noninformative censoring, which may not hold in all applications. Although the theoretical formulation accommodates both within-type and cross-type dynamic history effects, the applied analysis restricted the dynamics to within-type feedback to maintain computational tractability; cross-type effects were not estimated. Extensions could relax these assumptions, incorporate correlated or multilevel frailty structures, or allow richer dynamic interactions across event types. 

Despite these limitations, the proposed Bayesian framework provides a flexible and computationally efficient tool for jointly modeling multitype recurrent events and a terminal event. Its strong performance in simulations, stability under prior misspecification, and successful application to ALLHAT underscore its potential for broader use in complex event-history analyses.

\begin{spacing}{0.90}
\bibliographystyle{unsrt}  
\bibliography{sample} 
\end{spacing}
\begin{spacing}{1.0}
\begin{table}[H]
\centering
\caption{Simulation results for different combinations of sample size and frailty values under increasing failure rate ($\gamma^\ast = 1.1$).}
\label{tab:gamma1.1}

\renewcommand{\arraystretch}{1.2}
\setlength{\tabcolsep}{4pt}
\adjustbox{max width=\textwidth}{
\begin{tabular}{cccccc@{\hspace{13pt}}cccc}
\toprule
\multicolumn{2}{c}{\textbf{Parameter Combination}} & \multicolumn{4}{c}{\textbf{$\nu$ = 2}} & \multicolumn{4}{c}{\textbf{$\nu = 4$}} \\
\cmidrule(r){1-2} \cmidrule(r){3-6} \cmidrule(l){7-10}
\textbf{Parameter} & \textbf{n} & \textbf{Bias} & \textbf{SD} & \textbf{RMSE} & \textbf{CP} & \textbf{Bias} & \textbf{SD} & \textbf{RMSE} & \textbf{CP} \\
\midrule
$\nu$ & 100 & -0.836 & 0.362 & 0.911 & 0.86 & -0.300 & 0.355 & 0.464 & 0.99 \\
$\nu$ & 200 & -1.208 & 0.357 & 1.259 & 0.62 & -0.516 & 0.395 & 0.650 & 0.99 \\
\midrule
$\beta_{11}$ & 100 & -0.052 & 0.133 & 0.143 & 0.88 & -0.037 & 0.118 & 0.124 & 0.92 \\
$\beta_{12}$ & 100 & -0.016 & 0.103 & 0.104 & 0.89 & -0.021 & 0.110 & 0.111 & 0.90 \\
$\beta_{21}$ & 100 & -0.056 & 0.128 & 0.139 & 0.89 & -0.044 & 0.120 & 0.128 & 0.90 \\
$\beta_{22}$ & 100 & -0.002 & 0.119 & 0.120 & 0.87 & 0.006 & 0.093 & 0.093 & 0.95 \\
$\beta_{31}$ & 100 & -0.059 & 0.129 & 0.142 & 0.85 & -0.049 & 0.108 & 0.119 & 0.91 \\
$\beta_{32}$ & 100 & 0.031 & 0.111 & 0.115 & 0.91 & 0.018 & 0.096 & 0.098 & 0.90 \\
$\beta_{01}$ & 100 & -0.086 & 0.118 & 0.146 & 0.83 & -0.062 & 0.106 & 0.123 & 0.91 \\
$\beta_{02}$ & 100 & 0.019 & 0.106 & 0.108 & 0.86 & 0.008 & 0.099 & 0.100 & 0.88 \\
$\beta_{11}$ & 200 & -0.058 & 0.112 & 0.126 & 0.78 & -0.039 & 0.091 & 0.099 & 0.86 \\
$\beta_{12}$ & 200 & -0.012 & 0.087 & 0.088 & 0.83 & -0.009 & 0.081 & 0.082 & 0.89 \\
$\beta_{21}$ & 200 & -0.063 & 0.104 & 0.122 & 0.80 & -0.041 & 0.094 & 0.103 & 0.84 \\
$\beta_{22}$ & 200 & 0.003 & 0.083 & 0.083 & 0.85 & -0.001 & 0.085 & 0.085 & 0.84 \\
$\beta_{31}$ & 200 & -0.066 & 0.104 & 0.123 & 0.76 & -0.046 & 0.085 & 0.097 & 0.83 \\
$\beta_{32}$ & 200 & 0.011 & 0.089 & 0.089 & 0.85 & 0.014 & 0.072 & 0.073 & 0.85 \\
$\beta_{01}$ & 200 & -0.077 & 0.089 & 0.118 & 0.78 & -0.063 & 0.084 & 0.105 & 0.82 \\
$\beta_{02}$ & 200 & 0.019 & 0.081 & 0.083 & 0.86 & 0.002 & 0.069 & 0.069 & 0.89 \\
\midrule
$\gamma_{1}$ & 100 & -0.103 & 0.147 & 0.180 & 0.96 & -0.070 & 0.116 & 0.135 & 0.98 \\
$\gamma_{2}$ & 100 & -0.102 & 0.116 & 0.155 & 0.97 & -0.078 & 0.099 & 0.126 & 0.99 \\
$\gamma_{3}$ & 100 & -0.129 & 0.105 & 0.167 & 0.95 & -0.083 & 0.089 & 0.122 & 0.98 \\
$\gamma_{1}$ & 200 & -0.075 & 0.103 & 0.127 & 0.97 & -0.033 & 0.084 & 0.089 & 0.97 \\
$\gamma_{2}$ & 200 & -0.080 & 0.089 & 0.120 & 0.96 & -0.054 & 0.083 & 0.099 & 0.97 \\
$\gamma_{3}$ & 200 & -0.089 & 0.076 & 0.117 & 0.94 & -0.061 & 0.075 & 0.096 & 0.93 \\
\midrule
$\alpha_{1}$ & 100 & 0.020 & 0.101 & 0.103 & 0.98 & 0.060 & 0.089 & 0.107 & 0.98 \\
$\alpha_{2}$ & 100 & -0.008 & 0.101 & 0.101 & 0.99 & 0.029 & 0.087 & 0.092 & 0.99 \\
$\alpha_{3}$ & 100 & -0.023 & 0.093 & 0.097 & 0.99 & 0.015 & 0.073 & 0.075 & 0.99 \\
$\alpha_{1}$ & 200 & -0.012 & 0.095 & 0.096 & 0.99 & -0.035 & 0.800 & 0.087 & 0.98 \\
$\alpha_{2}$ & 200 & -0.041 & 0.091 & 0.100 & 0.97 & 0.010 & 0.083 & 0.083 & 0.96 \\
$\alpha_{3}$ & 200 & -0.039 & 0.075 & 0.084 & 0.99 & 0.001 & 0.067 & 0.067 & 0.98 \\
\bottomrule
\end{tabular}
}
\end{table}

\begin{table}[H]
\centering
\caption{Simulation results for different combinations of sample size and frailty values under decreasing failure rate ($\gamma^\ast = 0.9$).}
\label{tab:gamma0.9}
\small
\renewcommand{\arraystretch}{1.2}
\setlength{\tabcolsep}{4pt}
\adjustbox{max width=\textwidth}{
\begin{tabular}{cccccc@{\hspace{13pt}}cccc}
\toprule
\multicolumn{2}{c}{\textbf{Parameter Combination}} & \multicolumn{4}{c}{\textbf{$\nu$ = 2}} & \multicolumn{4}{c}{\textbf{$\nu = 4$}} \\
\cmidrule(r){1-2} \cmidrule(r){3-6} \cmidrule(l){7-10}
\textbf{Parameter} & \textbf{n} & \textbf{Bias} & \textbf{SD} & \textbf{RMSE} & \textbf{CP} & \textbf{Bias} & \textbf{SD} & \textbf{RMSE} & \textbf{CP} \\
\midrule
$\nu$       & 100 & -0.773 & 0.398 & 0.870 & 0.83 &  0.506 & 0.373 & 0.628 & 0.98 \\
$\nu$       & 200 & -1.060 & 0.368 & 1.122 & 0.57 & -0.160 & 0.408 & 0.438 & 0.99 \\
\midrule
$\beta_{11}$ & 100 & -0.077 & 0.130 & 0.151 & 0.83 & -0.059 & 0.121 & 0.134 & 0.88 \\
$\beta_{12}$ & 100 &  0.020 & 0.120 & 0.122 & 0.85 &  0.012 & 0.118 & 0.118 & 0.87 \\
$\beta_{21}$ & 100 & -0.062 & 0.139 & 0.152 & 0.85 & -0.064 & 0.123 & 0.139 & 0.88 \\
$\beta_{22}$ & 100 &  0.019 & 0.114 & 0.115 & 0.87 &  0.013 & 0.100 & 0.101 & 0.89 \\
$\beta_{31}$ & 100 & -0.071 & 0.127 & 0.146 & 0.83 & -0.050 & 0.115 & 0.125 & 0.90 \\
$\beta_{32}$ & 100 &  0.017 & 0.116 & 0.117 & 0.85 &  0.007 & 0.104 & 0.104 & 0.89 \\
$\beta_{01}$ & 100 & -0.092 & 0.106 & 0.140 & 0.84 & -0.076 & 0.113 & 0.136 & 0.88 \\
$\beta_{02}$ & 100 &  0.010 & 0.103 & 0.104 & 0.90 & -0.004 & 0.096 & 0.096 & 0.90 \\
$\beta_{11}$ & 200 & -0.074 & 0.113 & 0.135 & 0.79 & -0.060 & 0.100 & 0.116 & 0.81 \\
$\beta_{12}$ & 200 &  0.013 & 0.085 & 0.086 & 0.87 &  0.001 & 0.074 & 0.074 & 0.88 \\
$\beta_{21}$ & 200 & -0.075 & 0.112 & 0.135 & 0.78 & -0.051 & 0.089 & 0.102 & 0.85 \\
$\beta_{22}$ & 200 &  0.016 & 0.088 & 0.090 & 0.84 &  0.014 & 0.078 & 0.079 & 0.91 \\
$\beta_{31}$ & 200 & -0.063 & 0.103 & 0.121 & 0.78 & -0.047 & 0.089 & 0.101 & 0.81 \\
$\beta_{32}$ & 200 &  0.016 & 0.096 & 0.098 & 0.78 &  0.009 & 0.078 & 0.079 & 0.86 \\
$\beta_{01}$ & 200 & -0.076 & 0.093 & 0.120 & 0.77 & -0.057 & 0.080 & 0.098 & 0.85 \\
$\beta_{02}$ & 200 &  0.013 & 0.076 & 0.077 & 0.92 &  0.006 & 0.069 & 0.070 & 0.88 \\
\midrule
$\gamma_{1}$ & 100 & -0.082 & 0.125 & 0.149 & 0.98 & -0.040 & 0.100 & 0.107 & 0.99 \\
$\gamma_{2}$ & 100 & -0.100 & 0.113 & 0.151 & 0.97 & -0.071 & 0.103 & 0.125 & 0.97 \\
$\gamma_{3}$ & 100 & -0.116 & 0.099 & 0.152 & 0.96 & -0.097 & 0.095 & 0.136 & 0.97 \\
$\gamma_{1}$ & 200 & -0.055 & 0.095 & 0.110 & 0.96 & -0.028 & 0.089 & 0.094 & 0.97 \\
$\gamma_{2}$ & 200 & -0.081 & 0.098 & 0.128 & 0.93 & -0.055 & 0.081 & 0.098 & 0.96 \\
$\gamma_{3}$ & 200 & -0.090 & 0.088 & 0.126 & 0.92 & -0.062 & 0.076 & 0.098 & 0.96 \\
\midrule
$\alpha_{1}$ & 100 &  0.059 & 0.100 & 0.116 & 0.96 &  0.091 & 0.080 & 0.121 & 0.95 \\
$\alpha_{2}$ & 100 &  0.027 & 0.099 & 0.103 & 0.95 &  0.053 & 0.082 & 0.098 & 0.96 \\
$\alpha_{3}$ & 100 &  0.019 & 0.076 & 0.078 & 0.99 &  0.042 & 0.059 & 0.072 & 0.99 \\
$\alpha_{1}$ & 200 &  0.032 & 0.091 & 0.096 & 0.98 &  0.077 & 0.078 & 0.109 & 0.94 \\
$\alpha_{2}$ & 200 &  0.004 & 0.084 & 0.084 & 0.97 &  0.038 & 0.082 & 0.090 & 0.95 \\
$\alpha_{3}$ & 200 & -0.004 & 0.070 & 0.072 & 1.00 &  0.034 & 0.060 & 0.069 & 0.96 \\
\bottomrule
\end{tabular}
}
\end{table}

\begin{table*}[!t]
\centering
\begin{threeparttable}
\caption{Convergence diagnostics (Gelman--Rubin $\hat{R}$ and effective sample size) for the dynamic joint frailty model.}
\label{tab:table3}
\small
\renewcommand{\arraystretch}{1.20}
\setlength{\tabcolsep}{4.5pt}
\begin{tabular*}{\textwidth}{@{\extracolsep\fill}lcccccccccc@{}}
\toprule
& \multicolumn{1}{c}{\textbf{Frailty}}
& \multicolumn{8}{c}{\textbf{Regression Coefficients}} \\
\cmidrule(lr){2-2}
\cmidrule(lr){3-10}
\textbf{Stat}
& $\nu$
& $\beta_{11}$ & $\beta_{12}$ & $\beta_{21}$ & $\beta_{22}$ & $\beta_{31}$ & $\beta_{32}$ & $\beta_{01}$ & $\beta_{02}$ \\
\midrule
$\hat{R}$
& 1.01
& 1.00 & 1.00 & 1.00 & 0.99 & 1.00 & 0.99 & 1.00 & 1.00 \\[2pt]
ESS
& 1166
& 2731 & 2593 & 2620 & 2422 & 2556 & 2490 & 2573 & 2383 \\[2pt]
ESS(\%)
& 29.2
& 68.3 & 64.8 & 65.5 & 60.6 & 63.9 & 62.2 & 64.3 & 59.6 \\
\midrule
& \multicolumn{6}{c}{\textbf{Dynamic Parameters}} & & & \\
\cmidrule(lr){2-7}
\textbf{Stat}
& $\alpha_{1}$ & $\alpha_{2}$ & $\alpha_{3}$ & $\gamma_{1}$ & $\gamma_{2}$ & $\gamma_{3}$ & & & \\
\midrule
$\hat{R}$
& 1.01 & 1.00 & 1.00 & 1.01 & 1.01 & 1.01 & & & \\[2pt]
ESS
& 2245 & 2398 & 2444 & 1888 & 1763 & 1603 & & & \\[2pt]
ESS(\%)
& 56.1 & 60.0 & 61.1 & 47.2 & 44.1 & 40.0 & & & \\
\bottomrule
\end{tabular*}
\begin{tablenotes}[flushleft]\footnotesize
\item[$\dagger$] Total post burn-in samples are $4{,}000$ across four chains. Diagnostics meeting $\hat{R}\le 1.01$ and ESS $\ge 10\%$ of the total draws indicate acceptable convergence.
\end{tablenotes}
\end{threeparttable}
\end{table*}

\begin{table}[H]
    \centering
    \caption{Posterior summaries and hazard ratios from the dynamic joint frailty model.}
    \label{tab:table4}
    \small
    \renewcommand{\arraystretch}{1.2}
    \scalebox{0.95}{
    \begin{tabular}{l l c c c c c}
        \toprule
        \textbf{Event} 
        & \textbf{Variable} 
        & \textbf{Estimate} 
        & \textbf{Standard Error} 
        & \textbf{Lower CI} 
        & \textbf{Upper CI} 
        & \textbf{Hazard Ratio} \\
        \midrule
        \multirow{4}{*}{ACE} 
        & Amlodipine                 & 0.650 & 0.048 & 0.556 & 0.740 & 1.92 \\
        & Lisinopril                 & 0.894 & 0.047 & 0.808 & 0.986 & 2.45 \\
        & Race (black)               & 1.435 & 0.052 & 1.321 & 1.529 & 4.20 \\
        & Age (months)$^{\dagger}$   & 1.536 & 0.035 & 1.464 & 1.600 & 4.65 \\
        \midrule
        \multirow{4}{*}{CCE} 
        & Amlodipine                 & 0.597 & 0.041 & 0.526 & 0.676 & 1.82 \\
        & Lisinopril                 & 0.747 & 0.041 & 0.666 & 0.826 & 2.11 \\
        & Race (black)               & 1.523 & 0.051 & 1.425 & 1.617 & 4.59 \\
        & Age (months)$^{\dagger}$   & 1.412 & 0.032 & 1.348 & 1.474 & 4.10 \\
        \midrule
        \multirow{4}{*}{Death} 
        & Amlodipine                 & 0.943 & 0.040 & 0.860 & 1.019 & 2.57 \\
        & Lisinopril                 & 1.140 & 0.039 & 1.065 & 1.219 & 3.13 \\
        & Race (black)               & 1.966 & 0.051 & 1.864 & 2.056 & 7.14 \\
        & Age (months)$^{\dagger}$   & 1.973 & 0.032 & 1.908 & 2.033 & 7.20 \\
        \midrule
        \multirow{5}{*}{}
        & Frailty variance ($\nu$)   & 0.124 & 0.003 & 0.120 & 0.129 & ---- \\
        & $\alpha_{\text{ACE}}$      & 0.897 & 0.112 & 0.706 & 1.141 & ---- \\
        & $\alpha_{\text{CCE}}$      & 1.816 & 0.082 & 1.653 & 1.970 & ---- \\
        & $\gamma_{\text{ACE}}$      & 0.121 & 0.044 & 0.051 & 0.226 & ---- \\
        & $\gamma_{\text{CCE}}$      & 0.003 & 0.003 & 0.001 & 0.012 & ---- \\
        \bottomrule
    \end{tabular}
    }
    
    \vspace{0.15cm}
    \begin{minipage}{0.9\linewidth}
    \footnotesize
    $^{\dagger}$Age was standardized to have mean 0 and variance 1.
    \end{minipage}
    
\end{table}
    
\end{spacing}

\end{document}